%% file: ms.tex
\documentclass[iop]{emulateapj}

\newcommand{\CI}{\ion{C}{1}}
\newcommand{\CII}{\ion{C}{2}}
\newcommand{\CIV}{\ion{C}{4}}
\newcommand{\eqw}{\ensuremath{W_{\lambda}}}
\newcommand{\Ha}{\ensuremath{{\rm H}\alpha}}
\newcommand{\HI}{\ion{H}{1}}
\newcommand{\hst}{{\sl HST}}
\newcommand{\kms}{\ensuremath{{\rm km\,s}^{-1}}}
\newcommand{\lya}{\ensuremath{{\rm Ly}\alpha}}
\newcommand{\OVI}{\ion{O}{6}}
\newcommand{\Rvir}{\ensuremath{R_{\rm vir}}}

\begin{document}

\title{On the Size and Mass of Photo-ionized Clouds in Extended Spiral Galaxy Halos }
\author{Julie D. Davis, Brian A. Keeney, Charles W. Danforth, and John T. Stocke}
\affil{Center for Astrophysics and Space Astronomy, Department of Astrophysical and Planetary Sciences, University of Colorado \\ 389 UCB, Boulder, CO 80309; julie.davis@colorado.edu}

\shorttitle{Photo-ionized Clouds in Galaxy Halos}
\shortauthors{Davis et~al.}

\begin{abstract}

The size and mass of two circum-galactic medium (CGM) clouds in the halo (impact parameter = 65~kpc) of a nearby late-type galaxy, MGC~-01-04-005 ($cz = 1865$~\kms), are investigated using a close triplet of QSO sight lines \citep[the ``LBQS Triplet'';][]{crighton10}. Far ultraviolet spectra obtained with the Cosmic Origins Spectrograph (COS) aboard the {\sl Hubble Space Telescope} (\hst) find two velocity components in \lya\ at $\sim1830$ and 1900~\kms\ in two of these sight lines, requiring minimum transverse cloud sizes of $\geq10$~kpc. A plausible, but not conclusive, detection of \CIV\ 1548~\AA\ absorption at the higher velocity in the third sight line suggests an even larger lower limit of $\geq23$~kpc for that cloud. Using various combinations of constraints, including photo-ionization modeling for one absorber, lower limits on masses of these two clouds of $\gtrsim10^6~M_{\Sun}$ are obtained. Ground-based imaging and long-slit spectroscopy  of MCG~-01-04-005 obtained at the Apache Point Observatory (APO) 3.5m telescope find it to be a relatively normal late-type galaxy with a current star formation rate (SFR) of $\sim0.01~M_{\Sun}\,{\rm yr}^{-1}$. {\sl Galaxy Evolution Explorer} ({\sl GALEX}) photometry finds a SFR only a few times higher over the last $10^8$~yrs. We conclude that the CGM clouds probed by these spectra are typical in being at impact parameters of 0.4--0.5\,\Rvir\ from a rather typical, non-starbursting late-type galaxy so that these size and mass results should be generic for this class. Therefore, at least some CGM clouds are exceptionally large and massive. 

\end{abstract}


\section{Introduction}
\label{intro}

In the last few years the advent of the high throughput far-UV (FUV) Cosmic Origins Spectrograph \citep[COS;][]{green12} on the {\sl Hubble Space Telescope} (\hst) has allowed the discovery of a massive and extensive circum-galactic medium (CGM; a.k.a. gaseous halo) around late-type galaxies \citep{prochaska11, tumlinson11, tumlinson13, stocke13, werk13, borthakur13, stocke14, werk14, bordoloi14, lehner15}. COS FUV spectroscopy of background, bright QSO targets finds \HI\ often accompanied by low- (e.g., \CII, \ion{Si}{2}, \ion{Si}{3}) and/or high-ion (e.g., \CIV, \ion{Si}{4}, and \OVI) metal absorption lines at the redshifts of foreground, typically late-type galaxies. The absorptions occur ubiquitously within approximately one projected virial radius \citep[\Rvir; see][for a detailed discussion]{shull14} implying high covering factor \citep*{prochaska11, tumlinson11, stocke13} for the CGM around late-type galaxies in the current epoch. 

Studies of low-$z$ galaxy halos have been made by targeting QSO/galaxy pairs using \hst/COS \citep*{tumlinson13, stocke13, bordoloi14}. Using COS science team Guaranteed Time Observations (GTO), \citet{stocke13} studied late-type galaxy halos from extreme dwarfs and low surface brightness galaxies to $\sim L^*$ spirals at impact parameters $\leq1.5\,\Rvir$ finding associated \HI\ \lya\ absorption in all cases. The COS-Halos research team \citep*{tumlinson11, tumlinson13, werk13, bordoloi14} studied both late-type, star forming galaxies and early type galaxies with very low amounts of current star formation. The first COS-Halos studies were of $L>L^*$ galaxies. The later work of \citet{bordoloi14} concentrated on $L=0.1$--$1\,L^*$ late-type galaxies along with a few lower luminosity dwarfs. The COS-Halos program targeted sight lines whose impact parameters are $\leq0.5\,\Rvir$. 

These new \hst/COS observations were supplemented by the use of archival Space Telescope Imaging Spectrograph (STIS) FUV spectroscopy for a large number of very bright, low-$z$ QSOs in conjunction with ground-based spectroscopic surveys of galaxies near these sight lines \citep*{morris93, bowen97, tripp98, penton02, penton04, stocke06, prochaska11, stocke13}. Differing from the targeted surveys mentioned in the previous paragraph, these ``serendipitous'' studies found absorption systems first, then identified associated galaxies in some cases using the ground-based galaxy redshift surveys. In these ``serendipitous'' surveys approximately 50\% of absorbers with log N$_{HI}$ (cm$^{-2}$) $\geq$ 14.0 are found within the virial radius of an associated galaxy. This percentage decreases with decreasing column density. 

These two types of studies \citep[summarized in][]{stocke13} show \HI\ \lya\ absorption at comparable redshifts to foreground $L \geq 0.1\,L^*$ late-type galaxies at near unity covering factor out to 1--2\,\Rvir. Dwarfs also have significant associated absorption but at lesser ($\sim$ 50\%) covering factors within 1--2\,\Rvir. At all luminosities, these covering factors decline quite slowly out to several virial radii making it unclear where the CGM ends and the IGM begins if only \lya\ is considered. However, it has been known for some time \citep[e.g.,][]{chen01, stocke06} that metal absorption (\CIV\ and \OVI\ in particular) truncates rather dramatically at $\sim$ 0.3--1~Mpc from the nearest bright galaxy. It is plausible but not proven that metal-enriched gas at $Z \gtrsim 0.1~Z_{\Sun}$ metallicity occurs only within the confines of spiral-rich galaxy groups \citep{stocke14}, which have similar physical extents to this enriched gaseous reservoir. The detection of extensive CGMs around early-type galaxies remains controversial \citep{thom12}.

From the initial COS-Halos program results \citep{tumlinson11} it was already clear that these gaseous halos are very massive. However, determining a more accurate mass in photo-ionized halo clouds requires some simple modeling to determine the ionization state, density, and extent of these clouds. This standard photo-ionization modeling has been accomplished by two independent groups who find quite similar but not identical results. Both analyses assume single phase clouds in photo-ionization equilibrium with the extragalactic ionizing radiation, whose amount is fixed assuming the value from \citet{haardt12}. If a UV background intensity somewhat stronger than that of \citet{haardt12} is used \citep*[see][for a justification for this adjustment]{shull15}, the cloud densities and total masses determined would also increase by a similar factor. By determining the ionization parameter using the line ratios of metal ions in different ionization states, total cloud densities and indicative sizes can be calculated. Cloud masses follow by assuming quasi-spheroidal clouds. \citet{stocke13} analyzed 25 systems with at least two adjacent ion states, finding cloud densities and line-of-sight sizes of $n_H = 10^{-3}$ to $10^{-4}~{\rm cm}^{-3}$ and 0.1--20~kpc, respectively. \citet{werk14} analyzed the metal absorption systems in 44 clouds finding somewhat lower densities ($\langle n_{\rm H} \rangle = 10^{-4.2}~{\rm cm}^{-3}$) and larger cloud sizes. However, based on the limited number of input spectral line strengths (and their substantial uncertainties) as well as the unlikely assumption that the clouds are homogenous, single-phase gas, the results of these photo-ionization models are quite uncertain. Even in ideal circumstances these models provide only a substantial range for values of cloud density and size (and thus mass). Therefore, other cloud size constraints are required to be able to conclude more definitively that photoionized CGM clouds are large and massive.  

From these basic results the inferences derived for the mass of the CGM around a typical $L \geq 0.1\,L^*$ late-type galaxy diverge somewhat between these two studies. \citet{werk14} advocate for a single-phase CGM with density decreasing only very slowly away from the associated galaxy and $M_{\rm CGM} = 6.5\times10^{10}~M_{\Sun}$. \citet{stocke13} use their 25 observed clouds to construct an ensemble mass ($\sim 10^{10}~M_{\Sun}$) and a volume filling factor of 3--10\%. The latter results yield a significantly lower total CGM mass, leaving a large volume in the CGM unfilled by photo-ionized gas clouds.

\begin{figure*}[!t]
\epsscale{1.00} \centering \plotone{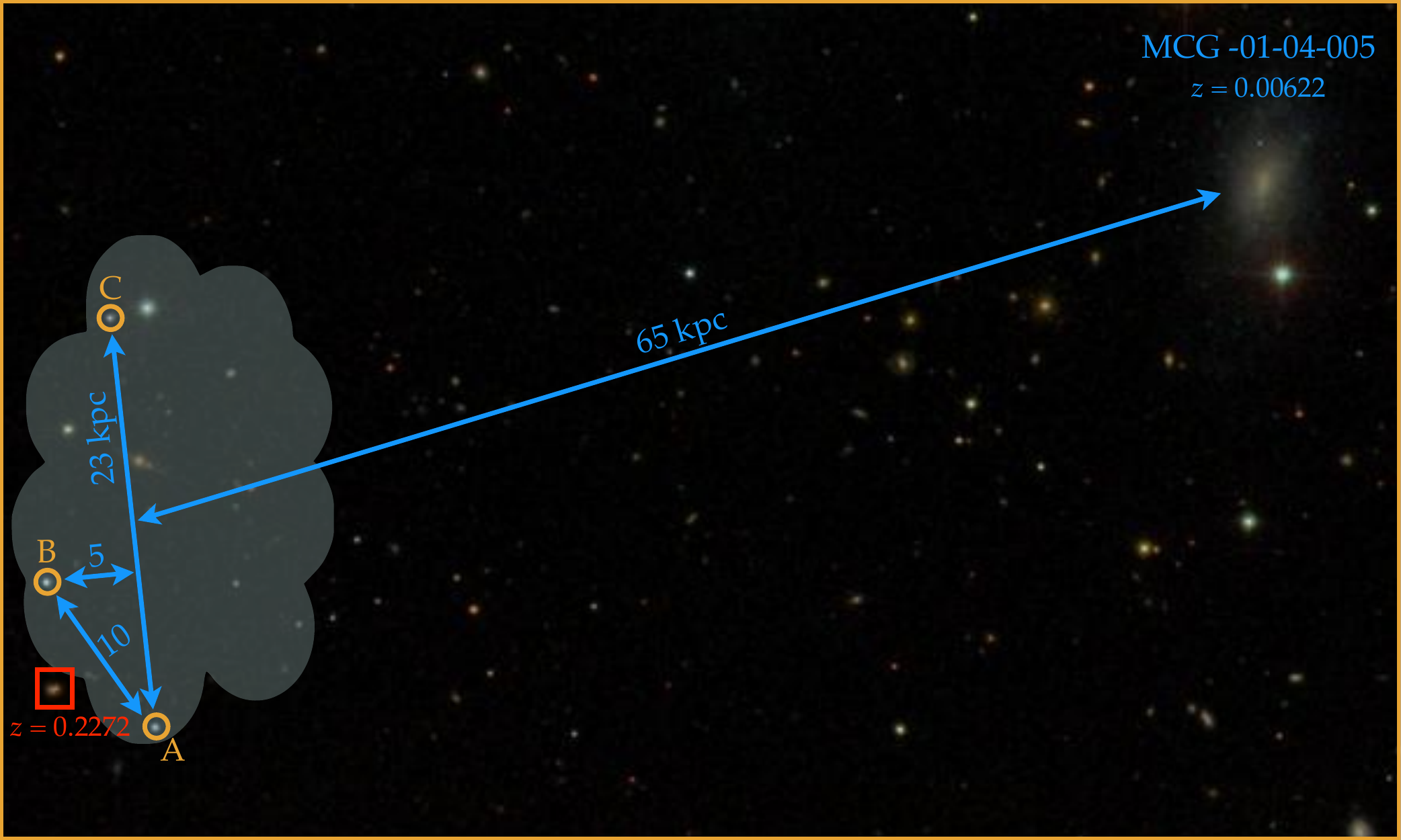}
\caption{An SDSS finder chart of the region surrounding the three QSO sight lines studied here (see Section~\ref{spectra}) on which we have overlaid {\it an illustrative cartoon} of the CGM cloud's extent if detected in all three sight lines. The QSO positions are indicated by the labels A--C, corresponding to the sight line labeling in the text. An indicative physical impact parameter for the QSO sight lines from MCG~-01-04-005 (Section~\ref{galaxy}) is also shown, as well as the distance between the QSO sight lines evaluated at $z=0.00622$. Finally, the position of the $z=0.2272$ galaxy identified by \citet [][; see Section~\ref{spectra}]{crighton10} is indicated by a red square. The image field of view is $10\arcmin\times6\arcmin$ and its orientation is north up, east left. All unlabeled distances are kiloparsecs.
\label{fig:context}}
\end{figure*}

Whether the CGM is completely filled with photo-ionized gas or not makes a difference for the interpretation of the detected \OVI\ absorption. In the \citet{werk14} analysis the \OVI\ must arise in gas co-mingled with the gas giving rise to the photo-ionized absorption regardless of whether the \OVI\ itself is photo-ionized or collisionally ionized. In the \citet{stocke13} picture the photo-ionized clouds are surrounded by a hotter medium which fills 90--97\% of the CGM volume; the interface between the clouds and a hot substrate create the collisionally ionized \OVI\ through shocks. \citet{stocke14} claim that broad, shallow \OVI\ absorption discovered in high-S/N COS spectra obtained by the GTO team \citep{savage14} is evidence for this hotter substrate. Unlike the stronger, narrower \OVI\ detected and studied by the COS-Halos Team, this shallower absorption suggests temperatures in the $T = 10^5$ to $10^{6.5}$~K range. An intra-group gas at these temperatures was predicted to be present in spiral-rich groups by \citet{mulchaey96}, analagous to the hot gas seen in rich clusters and elliptical-dominated rich groups of galaxies \citep{mulchaey00}. 

New constraints on cloud sizes can distinguish between these two pictures of the CGM, indirectly arguing for or against a very massive, hot, diffuse group gas surrounding these late-type systems. More generally, since the mass of a quasi-spherical cloud scales as cloud size $r^3$ while the total area of the cloud goes as $r^2$, the larger the individual halos clouds, the greater their mass, scaling as $r^3/r^2$ {\it in the limit of unity covering factor}.

In this paper we place constraints on the size of two halo clouds found around a rather normal $\sim0.1\,L^*$ galaxy using a triplet of sight lines closely spaced on the sky (10--23~kpc separation at the galaxy's Hubble distance of 27~Mpc; see Figure~\ref{fig:context}). These three closely-spaced QSOs were found in the Large Bright QSO Survey \citep[LBQS;][]{foltz87}, designated LBQS~0107--025A,B and LBQS~0107--0232 (referred to collectively as the ``LBQS triplet'' hereafter) and described in detail in \citet{crighton10}. In Section~\ref{spectra} we describe the \hst/COS spectra, concentrating on absorptions found in the three sight lines at $z=0.00622$. In Section~\ref{galaxy} we describe our new observations of galaxy MCG~-01-04-005 which bear on its recent star formation history. The discussion in Section~\ref{discussion} focuses on the constraints on cloud size obtained using these observations. In Section~\ref{conclusion} we summarize our conclusions.

\section{\hst/COS FUV Spectroscopy of the LBQS Triplet}
\label{spectra}

\citet{crighton10} use the absorption line data from the \hst/COS FUV spectra of the LBQS triplet to constrain the relationship between galaxies and absorbers at $z<1$, finding an excess of groups of
galaxies compared to expectations. A high column density Lyman limit system (LLS) is present in LBQS~0107--0232 (J0107C hereafter) at $z=0.557$ and a metal-enriched absorber is found in LBQS~0107--025A,B (J0107A and J0107B hereafter, respectively) at $z=0.227$, located 200~kpc from a bright galaxy at the same redshift (see Figure~\ref{fig:context}). \citet{muzahid14} used the common \OVI\ absorptions in this system to infer that the \OVI-absorbing gas is very large on the sky (600--800~kpc) and must therefore contain a substantial amount of mass ($\geq 10^{11}~M_{\Sun}$). This inferred mass is similar to the amount inferred by \cite{stocke13} to be in the hot CGM phase found by \citet{savage14} and  \citet{stocke14}. \citet{muzahid14} also analyzed the photo-ionized phase in the $z=$0.227 absorber, finding evidence that this cooler cloud is quite small ($<1$~kpc) suggesting that it is a high velocity cloud imbedded in the hot \OVI-absorbing substrate.

The same COS spectra used by \citet{crighton10} and \citet{muzahid14} were used by this project. All targets were observed with the Cosmic Origins Spectrograph (COS) as part of HST program 11585 (PI: Crighton) in late 2010 and early 2011.  J0107A and J0107B were observed with both far-UV, medium resolution gratings COS/G130M ($1135<\lambda<1450$ \AA) and COS/G160M ($1400<\lambda<1795$ \AA).  J0107C was observed in the COS/G160M grating only due to the LLS discussed above.  Observations made at several different grating positions give continuous spectral coverage over the range $1135<\lambda<1795$ \AA\ ($1400<\lambda<1795$ \AA\ for J0107C) at an approximate resolution of $\mathcal{R}=\lambda/\Delta\lambda\approx18,000$ ($\Delta v\approx 17$~\kms).  

The calibrated, one-dimensional spectra for each exposure were obtained from the Mikulski Archive for Space Telescopes (MAST).  The individual exposures were then coadded using standard IDL procedures described in detail by \citet{danforth10}.  Briefly, the individual exposures were binned by three pixels, cross-correlated around strong Galactic absorption features, interpolated onto a common wavelength scale, and combined using an exposure-weighted coaddition scheme.  The combined spectra shows the expected smooth continuum and narrow absorption features.  The data quality varies over the spectral range due to the different sensitivities and exposure times in the two detectors.  Total exposures times, median S/N per resolution element, and median flux levels for each grating are given in Table~\ref{tab:obs}.

\input tab1.tex

\subsection{Absorption Line Analysis}
\label{spectra:abs}

\begin{figure*}[!t]
\epsscale{1.00} \centering \plotone{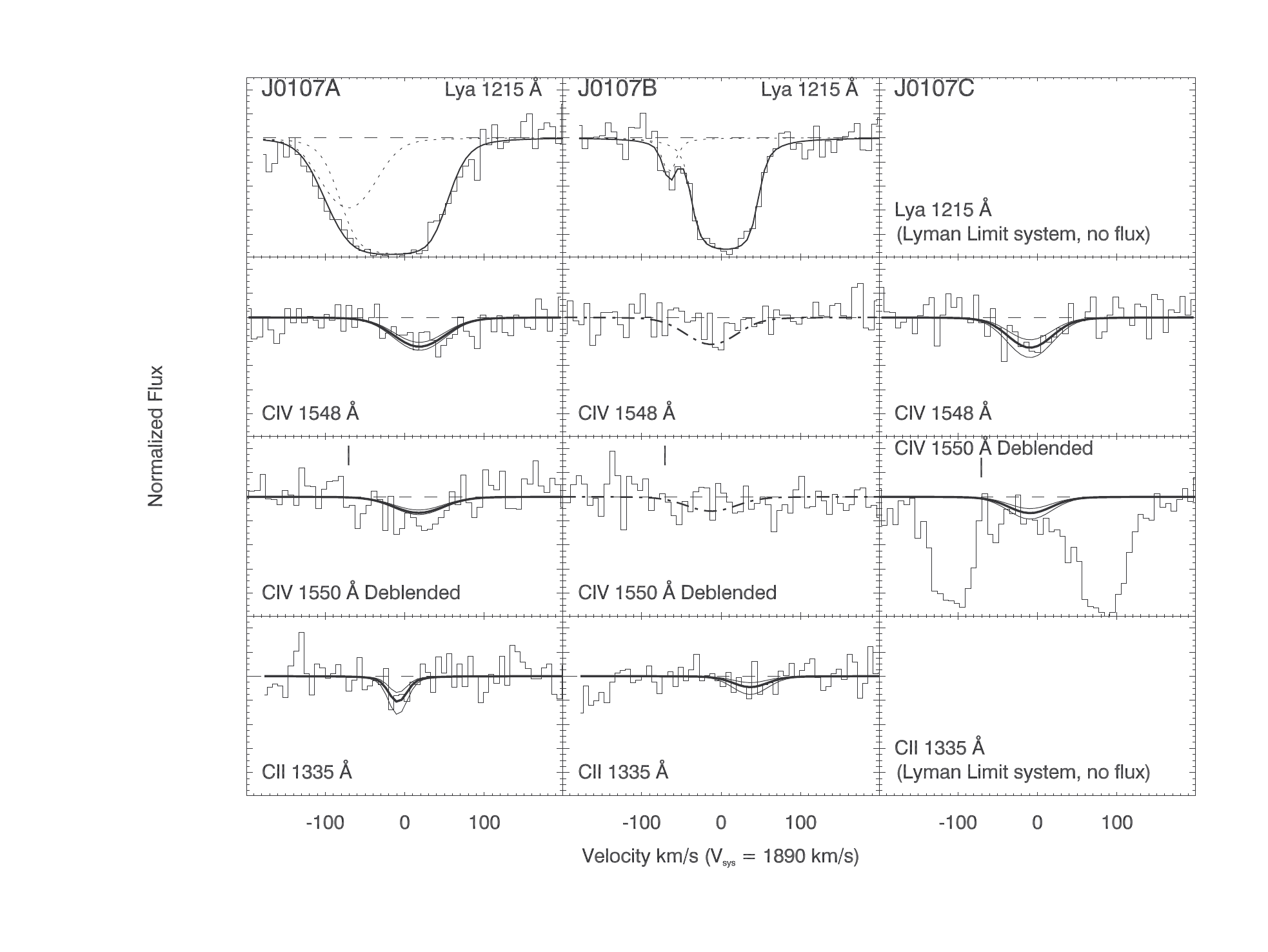}
\vspace{-4em}
\caption{\HI\ \lya\ (first row), \CIV\ doublet absorptions (2nd and 3rd rows) and C~II (4th row) in all three sight lines, over-plotted with profile fits from Table 2. The components of each \lya\ absorption are represented with dotted lines. In J0107A the modeled profile (thick line) with its errors (thin lines) are plotted for the strong \CIV\ doublet line. Direct fits to the weaker doublet line produce column values much too large for the expected doublet ratio, so it is likely that the line is blended with an unknown weak line. Thus a predicted profile based on the measured column density and b-values of the 1548 \AA~ line is plotted instead. The \CIV\ model fits in J0107B (dash-dot lines) represent upper limits centered on the wavelength calculated using J0107A's \lya\ redshift. The \CIV\ 1548 \AA~ line in J0107C is also the modeled profile with error bars. The 1550 \AA~ profile in J0107C is a predicted profile like what is displayed for the the J0107A weak C\,IV. In row \#3 the Galactic \CI\ line was removed using a profile generated from other Galactic \CI\ lines -- the entirety of the region of this subtraction is 1559.6 \AA~ to 1561.7 \AA~, corresponding to -200 to +200 \kms. The center of the removed \CI\ profile is noted with a vertical tick mark. Some residuals from the \CI\ deblending may remain in this region. The 4th row displays \CII\ 1335 \AA\ absorption for J0107A and J0107B with modeled profiles overplotted. 
\label{fig:absall}}
\end{figure*}

To establish the extent of the circumgalactic halo clouds around MCG -01-04-005 \citep [systemic radial velocity = 1864 $\pm$ 5 km s$^{-1}$;][]{koribalski04} we utilize the presence or absence of \HI\ \lya\ and \CIV\ 1548, 1550~\AA\ absorption at the host galaxy redshift in the LBQS triplet (see Figure~\ref{fig:absall} for all detected absorbers at the redshift of MGC -01-04-005). 

The \HI\ \lya\ absorptions in J0107A and J0107B are best-fit using two components (see Table 2). In velocity space, both sight lines include a component at $\sim1830$~\kms\ and one at $\sim1900$~\kms.  The specific values for these fits can be seen in Table 2. All of the line fits have physically reasonable $b$-values except the 1836 \kms\ component in J0107B, which is weak enough that a broader, physically-plausible, profile is possible within the statistical errors (see Figure~\ref{fig:absall}). Full line profiles based on the Voigt fit parameters in Table 2 are over-plotted as thick solid lines for all absorption features. For the Ly$\alpha$ profiles, the dotted lines indicate the Voigt fits for each component; for the \CIV\ profiles, the thin lines indicate the 1$\sigma$ uncertainty on these fits. Special care is required when inferring \HI\ column densities from Voigt profile fits to \lya\ profiles; we address these concerns in Section~4.1.

The wavelengths of \lya\ and \CII\ in J0107C are obscured by a strong, higher redshift LLS. However, in this case a reasonably strong \CIV\ absorption is visible at $\sim 6\sigma$ in the stronger of the two doublet lines (see Figure~\ref{fig:absall} third column and Table 2. The velocity of the \CIV\ in J0107C corresponds to within the errors with a comparably strong \CIV\ in J0107B consistent with the $\sim1900$~\kms\ \lya\ component. While no C~IV is detected in J0107B, the much stronger \lya\ absorber in J0107A compared to J0107B means that the $\sim1900$~\kms\ \lya\ component can be at a similar metallicity in all three sight lines. The strength of the \CIV\ in J0107C compared with J0107B suggests that a comparably strong \lya\ is present at that location, although undetectable due to the higher redshift LLS. Likewise, C~II 1335 \AA\ is weakly detected in J0107A \& B, consistent with belonging to the higher velocity absorber (see Figure 2).

\input all_species_table.tex

Due to intervening ISM and \lya\ forest lines, several of the \CIV\ candidate lines require de-blending to confirm their presence. At the galaxy's redshift, and thus in all three spectra, the weaker line of the \CIV\ doublet is blended with Galactic neutral carbon (\CI\ 1560~\AA) ISM absorption shown at a velocity of $-72$~\kms\ in Figure~\ref{fig:absall}. We modeled out the neutral Galactic carbon by simultaneously Voigt-fitting other \CI\ transitions in the same spectrum to obtain a \CI\ column density. A model \CI\ profile using the parameters obtained was then subtracted from the blended region of interest. This model has already been removed from the spectra presented in Figure~\ref{fig:absall}. The location of the \CI\ wavelength is noted in the third row of Figure~\ref{fig:absall}.

In J0107B, only a $3\sigma$ upper limit on \CIV\ is reported after the \CI\ model was removed. For J0107A \& J0107C, we were able to model cleanly the \CIV\ 1548~\AA\ line (a $6\sigma$ detection). Particularly for J0107A (see Figure~\ref{fig:absall}) the 1548~\AA-derived model seems to fit the de-blended 1550 \AA\ feature, implying that the de-blending process was reasonably successful with only small \CI\ residuals. However, this line more than likely contains additional contaminants besides \CI\, making its parameters less well-determined; i.e., the 1550 \AA\ detection appears too strong for the observed 1548 \AA\ line strength. Table 2 thus lists proxy values for the \CIV\ 1550 \AA\ feature in J0107A based on the values obtained from the 1548~\AA\ measurement.

In J0107C only the stronger \CIV\ line is well-measured at 1560.9~\AA\ due to the presence of \lya\ forest lines at other redshifts in the region of the weaker doublet line. Nevertheless, after the \CI\ de-blending, a weak \CIV\ 1550~\AA\ feature does seem to be present between two strong \lya\ absorbers. This tentative detection of the weaker \CIV\ doublet line in J0107C is suggestive but not conclusive due to the higher redshift \lya\ lines in its vicinity.

\begin{figure}[!t]
\epsscale{1.00} \centering \plotone{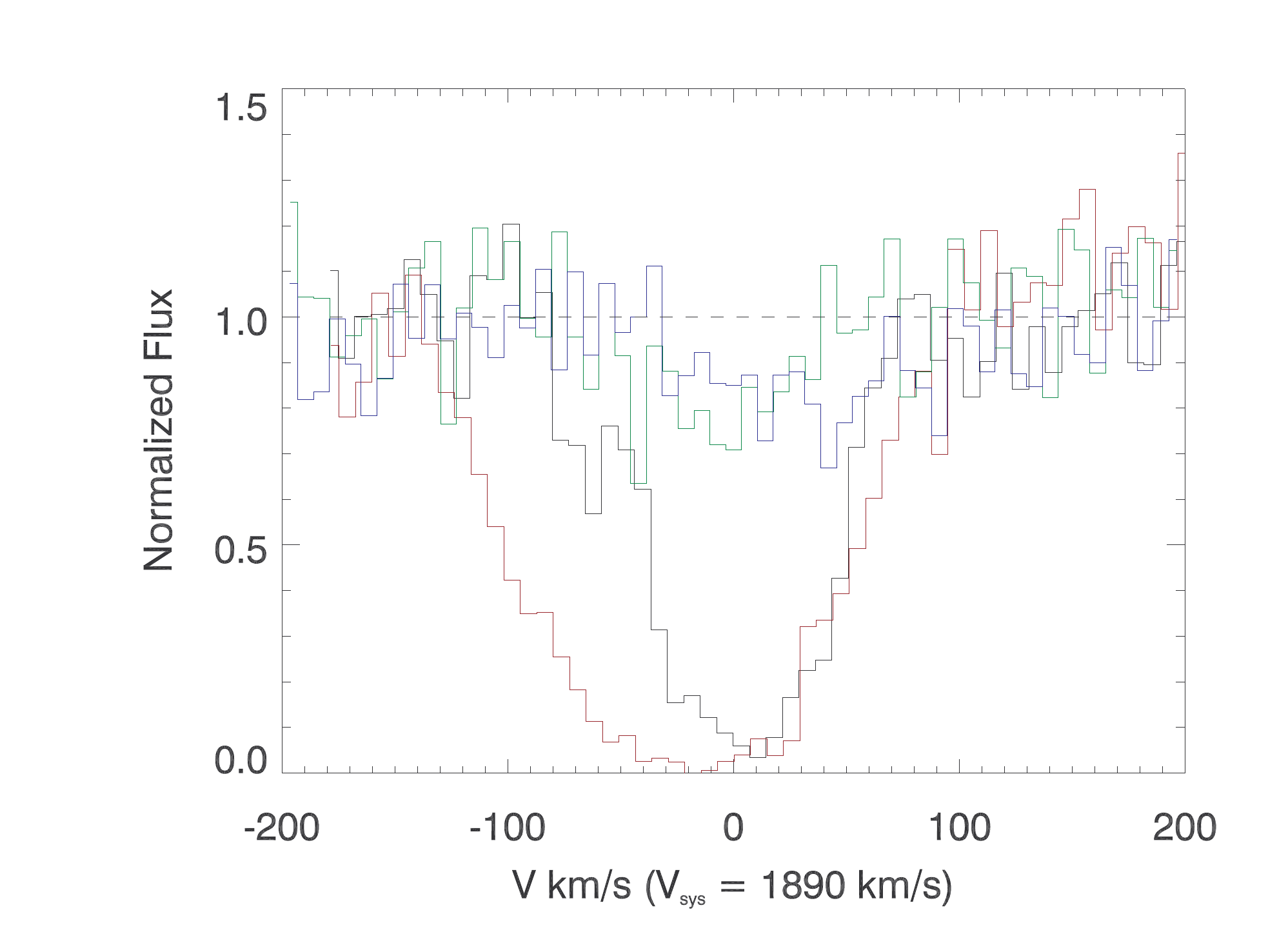}
\vspace{-1em}
\caption{Relative velocities of the absorption systems, using \lya\ in J0107A (red) and J0107B (black), and \CIV\ 1548~\AA\ in J0107A (blue) and J0107C (green). The \CIV\ absorptions in J0107A \& C have very similar profiles and velocity extents and are associated entirely with the higher velocity \lya\ absorption system.
\label{fig:absvel}}
\end{figure}

With no other definite lines than the \CIV\ 1548~\AA\ identification at the same redshift in J0107C, we must entertain the possibility that this line is actually a spurious, weak feature at a different redshift. \lya\ absorbers are by far the most common in the IGM so the alternative interpretation of the 1560.9~\AA\ feature as \lya\ at $z=0.284$ is suggested.  The line has an observed equivalent width of $\eqw=96\pm44$~m\AA\ with a fitted $b=34\pm11$~\kms.  Interpreted as a weak \lya\ absorber on the linear part of the curve of growth, this corresponds to a rest equivalent width of $74\pm34$~m\AA, or $\log{N_{\rm H\,I}}=13.2\pm0.2~{\rm cm}^{-2}$. \citet{danforth15} find a bivariate distribution of \HI\ absorbers with $d^2{\cal N}(\log{N})/dz\,d\log{N} = 110\pm6$ at this column density. Multiplying by a column density bin width equal to the full range of the measurement uncertainty (0.4~dex), gives $d{\cal N}/dz=44\pm2$, the frequency of \HI\ absorbers per unit redshift with a column density $\log{N}=13.2\pm0.2$.
The 1560.9 \AA\ line is at the right wavelength to be \CIV\ associated with MGG -01-04-005, but we don't know the precise velocity of the \HI\ absorption in this particular sight line (see Figure 2).  Instead, we estimate the velocity range over which it might be expected using the \lya\ and \CIV\ absorption from the other two sight lines: $cz= 1860 \pm 50$~\kms.  A full-width of 100~\kms\ corresponds to $\Delta z=3.3\times10^{-4}$.  Multiplying this $\Delta z$ by the \lya\ absorber frequency at that column density and redshift, the probability for finding an unrelated \lya\ line by chance at the location of the expected \CIV\ line is $\sim$1.5\%.  This probability scales linearly with the redshift width $\Delta z$ of the allowed line centroid.

In summary the \lya\ and \CIV\ absorption lines in these three sight lines are over-plotted in Figure 3. The \lya\ lines in J0107A \& J0107B share both velocity components although the lower velocity component is quite weak in J0107B. J0107A \& J0107C have \CIV\ absorption which aligns with the higher velocity component and with each other. The velocity and velocity width coincidences between these two features argue in favor of hypothesis that the J0107C feature is also \CIV~. While the \CIV\ 1548~\AA\ identifications are confirmed by the presence of weak 1550~\AA\ absorption in J0107A \& B, the \CIV\ 1548~\AA\ identification in J0107C could instead be an intervening \lya\ absorber at a different redshift. Therefore, the higher velocity component is shared by at least two and maybe all three sight lines. The lower velocity component is present in J0107A without doubt and probably present, although weakly, in J0107B. While there is no evidence for the lower redshift component in J0107C, we can not rule out its presence since we have only \CIV\ coverage in that sight line and the lower redshift component appears to be metal-free at the sensitivity level of the COS spectroscopy in-hand.

\section{Foreground Galaxy: MCG~-01-04-005}
\label{galaxy}

The galaxy MCG~-01-04-005 lies to the WNW of the LBQS triplet and is projected $8\farcm8$ (68~kpc) on the sky from J0107A, $9\farcm1$ (71~kpc) from J0107B, and $8\farcm3$ (64~kpc) from J0107C (see Figure~\ref{fig:context}). Its \HI\ 21-cm centroid velocity \citep[$1865\pm6$~\kms;][]{meyer04} is in-between the two velocity components detected in FUV absorption (see Section~\ref{spectra:abs}) and it has a $B$-band luminosity of $\sim0.07\,L^*$ \citep{doyle05}.

\subsection{Galaxy Imaging and SFR}
\label{galaxy:img}

\begin{figure}[!t]
\epsscale{1.00} \centering \plotone{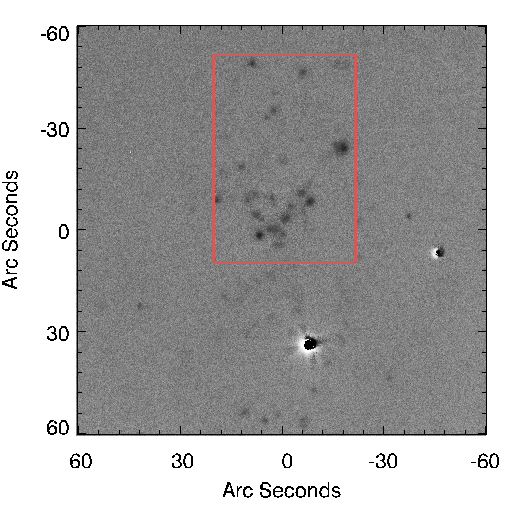}
\caption{Continuum-subtracted narrowband H$\alpha$ image of MCG~-01-04-005. The image has a $2\arcmin\times2\arcmin$ field-of-view and is oriented north up, east left. The rectangular region from which the galaxy's \Ha\ flux was extracted is indicated.Two bright stars in the field have positive and negative residuals due to changes in the seeing between the expsoures.
\label{fig:halpha}}
\end{figure}

\Ha\ and R-band images of MCG~-01-04-005 were obtained in November 2013 using the SPIcam imager on the Apache Point Observatory 3.5m telescope (APO). Six 300 sec exposures in the R-band and six 900 sec exposures in \Ha\ (R: 6492~\AA, $\Delta\lambda$ = 1544~\AA; \Ha: 6570~\AA, $\Delta\lambda$ = 75~\AA) were acquired under $\sim1\arcsec$ seeing conditions. The narrowband filter covers an area smaller than the detector field-of-view (FOV), yielding an unvignetted FOV of $\sim3\farcm5\times3\farcm5$, from which the subimage in Figure~\ref{fig:halpha} is extracted. A standard photometric procedure of subtracting a scaled R-band image from the \Ha\ image was used to isolate \Ha\ emission. In the pure emission-line image, counts from a $42\arcsec\times62\arcsec$ ($150\times220$~pixel; see Figure~\ref{fig:halpha}) rectangular area bounding the estimated extent of the galaxy emission were measured. An equal-sized rectangular sky region from the same image was subtracted. Assuming appropriate background subtraction and Poissonian error statistics, the resulting galaxy \Ha\ count rate was $54\pm1~{\rm cts\,s^{-1}}$. However, our \Ha\ count rate is contaminated by [\ion{N}{2}] emission that closely flanks \Ha\ and thus falls within the filter bandpass. To remove this, optical spectra of MCG~-01-04-005 were obtained on the same night with the Dual Imaging Spectrograph (DIS) at APO. The two [\ion{N}{2}] emission features flanking the \Ha\ emission peak were measured and the ratio of emission was found to be $F({\rm N\,II})/F(\Ha) = 0.31$. Therefore our corrected pure \Ha\ count rate is $37\pm1~{\rm cts\,s^{-1}}$.  

Using the standard star BD+28 4211 to flux calibrate, an observed \Ha\ flux of $(2.09\pm0.06) \times10^{-14}~{\rm ergs\,s^{-1}\,cm^{-2}}$ was measured for the galaxy. Galactic foreground extinction was determined using the extinction law of \citet{fitzpatrick99} assuming $R_V=3.1$ and $E(B-V) = 0.043\pm0.006$~mag \citep{schafly11}, yielding an extinction-corrected \Ha\ flux of $(2.29 \pm0.07)\times10^{-14}~{\rm ergs\,s^{-1}\,cm^{-2}}$; no extinction intrinsic to MCG~-01-04-005 was assumed. We use a luminosity distance of $\sim27$~Mpc \citep{bennet13} (also consistent with \citet{tully08} distance measurements to an equivalently redshifted galaxy group). This corresponds to a luminosity of $L_{\Ha} = (1.99\pm0.06)\times10^{39}~{\rm ergs\,s^{-1}}$. Using the calibration of \citet*{hunter10} we obtain a current SFR of $\approx 0.01~M_{\Sun}\,{\rm yr}^{-1}$. 

We can also infer a galaxy's recent SFR from its FUV luminosity. MCG~-01-04-005 was observed by the {\sl Galaxy Evolution Explorer} ({\sl GALEX}) in both the FUV and NUV imaging bands as part of its All Sky Imaging Survey. These shallow (108 sec) exposures found a {\sl GALEX} FUV magnitude of $17.31\pm0.08$ for MCG~-01-04-005 \citep[we have incorporated the absolute photometric uncertainty of {\sl GALEX} into this magnitude uncertainty;][]{morrissey07}, which corresponds to an extinction-corrected luminosity of $L_{\rm FUV} = (5.1\pm0.4)\times10^{26}~{\rm ergs\,s^{-1}\,Hz^{-1}}$ using the procedure outlined above. This luminosity implies that the galaxy's SFR is $\approx 0.06~M_{\Sun}\,{\rm yr}^{-1}$ \citep{hunter10}, $\sim6$ times higher than the \Ha-derived value. 

The higher SFR derived from the {\sl GALEX} FUV luminosity is not surprising since \Ha\ measures star formation over the past 10~Myr, whereas FUV luminosity measures star formation in the past 10--100~Myr \citep{hunter10}. Nevertheless, as evidenced by its low \Ha- and FUV-inferred SFRs, MCG~-01-04-005 is not currently a starburst galaxy nor has it been one in the recent past. All indications are that it is a rather normal $L \sim 0.1\,L^*$ galaxy.

\section{Size of a Typical Photo-ionized Halo Cloud}
\label{discussion}

These three sight lines present the opportunity to obtain limits on the physical extent of two typical late-type galaxy halo clouds. From the aspect of its SFR, MCG~-01-04-005 appears to be entirely normal, certainly not a starburst, over the recent past. The impact parameter for these sight lines is a modest number of disk radii and only 0.4--0.5\,\Rvir\ \citep[based on a halo-matching formalism;][]{stocke13}. This is comparable to or less than the 0.5--1\,\Rvir\ region for which very high covering factors in \lya\ and/or \OVI\ are found around late-type galaxies \citep{tumlinson11, stocke13, bordoloi14}. Therefore, this situation appears entirely typical for the study of photo-ionized halo clouds around late-type galaxies.

There are two velocity components to the \lya\ absorption, which are present in two sight lines: J0107A and J0107B, separated by 10~kpc at the distance of MCG~-01-04-005. This match sets firm lower limits on the transverse sizes on the sky of these two clouds. While Ly$\alpha$ is quite strong in both sight lines for the $\sim1900$~\kms\ absorber, the lower velocity ($\sim1830$~\kms) \lya\ absorption in J0107B is so much weaker than in J0107A that the lower velocity cloud may be only slightly $>10$~kpc in extent on the sky. These firm lower limits are already larger than most of the CGM cloud sizes inferred by \citet{stocke13} and near the median value for the sizes inferred by \citet{werk14}. Assuming a typical CGM cloud total hydrogen density of $n_{\rm H}=10^{-3.5}~{\rm cm}^{-3}$ for a spherical CGM cloud yields a lower limit on the mass of these two clouds: $\gtrsim4.5\times10^6~M_{\Sun}$. However, additional information is available that bears on the sizes and masses of these clouds.   

For the J0107C sight line, the absorption due to these clouds is ambiguous due to the presence of a LLS at higher redshift which obscures the portion of the UV spectrum which contains \lya\ associated with MCG~-01-04-005. However, in this case there is a shallow absorption present at just the right velocity to be \CIV\ 1548~\AA\ in the $\sim1900$~\kms\ absorber. Additionally, this feature has a similar line width to the \CIV\ found in the J0107A sight line (see Figure~\ref{fig:absvel}), so that \CIV\ is a very plausible identification for this feature. However, we cannot rule out entirely that this absorption line could be \lya\ at a higher redshift, although the probability of such a chance coincidence is small ($<2$\%). Additionally, there is a marginal detection of \CIV\ 1550~\AA\ in J01017C which falls between two strong, intervening \lya\ absorbers. Therefore, only slightly more speculatively, we conclude that the higher velocity ($\sim1900$~\kms) cloud extends $>23$~kpc on the sky. While we can use this limit as the minimum cloud extent in one dimension (approximately tangential to the impact parameter vector; see Figure 1), we do not know the other dimensions explictly. A plausible lower size limit for the other dimension on the plane of the sky can be set at $\geq5$~kpc using the distance from the J0107B sight line to the cord connecting sight lines A and C (since our observations suggest a continuous absorption between sight lines A \& C; see Figure 1). Assuming an ellipsoidal cloud with axes of $\mathrm{5~kpc\times5~kpc\times23~kpc}$ finds a minimum mass of $\gtrsim5\times10^6~M_{\Sun}$. Assuming $>23$~kpc for all three dimensions of this cloud, results in nearly a 10$^8~M_{\Sun}$ lower limit.

\subsection{Uncertainties in \HI\ Column Density}

In Section~4.2 we will use {\it CLOUDY} \citep{ferland98} photo-ionization models to attempt to constrain the line-of-sight thickness of the $\sim1900$~\kms\ absorber in J0107A. {\it CLOUDY} outputs rely on accurate absorption-line measurements, but the $N_{\rm H\,I}$ value from our Voigt profile fits requires re-evaluation as the \lya\ profile is slightly saturated in both velocity components and spectroscopy of the higher-order Lyman lines is not available in this case. The values listed in Table 2 for the \lya\ line are the values that minimize the $\chi^2$ of the Voigt-profile fit, but since we are on the flat part of the curve of growth there is a very large uncertainty in column density and $b$-value for a given equivalent width. 

Thus, we will treat the \HI\ column density listed in Table~2 for the $\sim1900$~\kms\ absorber as untrustworthy and estimate its value indirectly. This is unfortunate but unavoidable when the only \HI\ line that can be fit is a saturated \lya\ profile with no damping wings. A famous precedent from the literature may help illustrate this point. \citet{weymann95} observed 3C~273 with \hst/GHRS  using the G160M grating and found a best-fit Voigt profile with $\log{N_{\rm H\,I}} = 14.22\pm0.07$ and $b=34\pm3$~\kms\ for the $\sim1585$~\kms\ \lya\ absorber associated with the Virgo cluster. We note that the S/N and spectral resolution of this \hst/GHRS spectrum of 3C~273 are comparable to those of the \hst/COS spectrum of J0107A and the best-fit parameters from the \lya\ fits are comparable as well. Later, {\sl Far-Ultraviolet Spectroscopic Explorer (FUSE)} data became available for 3C~273 and absorption from Ly$\beta$-Ly$\theta$ were found at the same redshift \citep{sembach01}. When a curve-of-growth was generated from these higher-order Lyman series lines \citet{sembach01} found $\log{N_{\rm H\,I}} = 15.85^{+0.10}_{-0.08}$ and $b=16\pm1$~\kms. The $b$-value was reduced by a factor of 2 compared to the initial \citet{weymann95} value and the column density increased by a factor of more than 40! \citet{sembach01} attributed the discrepancy partially to very low column density component structure that affects the \lya\ profile but is undetectable in the higher-order Lyman series lines. Therefore, we must treat an $N_{\rm H\,I}$ value obtained from Ly$\alpha$ profile fitting alone as uncertain, very likely a lower limit, unless other physical constraints can be identified.

Since $N_{\rm H\,I}$ is underconstrained by the COS data alone, we have imposed the condition that the J0107A absorber metallicity should be consistent with the galaxy metallicity. This is reasonable because there are no other bright galaxies close to MCG~-01-04-005; the nearest brighter galaxy is NGC~448, an $0.6\,L^*$ S0 galaxy $\sim750$~kpc away on the sky. This distance on the sky is nearly 5$\times$ the virial radius for NGC~448 and close to the maximum distance that metals have been detected from any bright galaxy in the current epoch \citep{stocke06, stocke13}. We conclude that any metals present in these absorbers have originated in MGC~-04-01-005. If gas expelled by MCG~-04-01-005 mixed with IGM gas it could have an even lower metallicity and an even higher $N_{\rm H\,I}$. However, such extreme values do not give Ly$\alpha$ profile fits consistent with the COS spectrum of J0107A nor do they satisfy the photo-ionization modeling requirements presented below. 

We have measured a metallicity for MCG~-01-04-005 of $[{\rm O/H}] = -0.15\pm0.19$ (i.e., $Z_{\rm gal}\sim0.7\,Z_{\Sun}$) using our APO/DIS long slit spectrum (see Section 3.1) with the N2 index calibration of \citet{pettini04} and the solar oxygen abundance of \citet{asplund09}. For the absorber metallicity to be consistent with the galaxy metallicity, $\log{(Z_{\rm abs}/Z_{\Sun})} = -0.53$ to 0.23 (i.e., within $2\sigma$ of the galaxy's nominal value of $[{\rm O/H}] = -0.15$, or the 95\% confidence band). To satisfy this constraint the \HI\ column density of the $\sim1900$~\kms\ absorber in J0107A must lie in the range $\log{N_{\rm H\,I}} \approx 15.5\pm0.5$. In the photo-ionization models that follow, this range of values can reproduce all of the observed metal-line columns and limits; i.e., for lower \HI\ columns the \CII\ detection implies a metallicity that is too high and for higher \HI\ columns the \ion{Si}{2} and \ion{Si}{4} upper limits imply a metallicity that is below this range.

\subsection{Photo-ionization Modeling}

In this sub-Section we use {\it CLOUDY} photo-ionization modeling both to obtain additional size constraints for the $\sim1900$~\kms\ absorber in J0107A and also to show that such size constraints are quite uncertain using the minimal data in-hand: $N_{\rm H\,I}$, $N_{\rm C\,II}$, $N_{\rm C\,IV}$, and limits on the column densities of \ion{Si}{2}, and \ion{Si}{4} (see Table 2). We use a plane-parallel {\it CLOUDY} model irradiated by the extragalactic ionizing field as specified by \citet{haardt12} to search for a single-phase solution that self-consistently reproduces the measured column densities and limits. We assume that all species are solely photo-ionized and examine the temperature, $T$, hydrogen density, $n_{\rm H}$, neutral fraction, $f_{\rm H\,I}$, and line-of-sight thickness, $t$, of the modeled cloud as a function of ionization parameter, $U=n_{\gamma}/n_{\rm H}$, and metallicity, $Z$. 

Since the \HI\ column density derived from our Voigt profile fits to \lya\ is very uncertain, we use the {\it CLOUDY} models to constrain $N_{\rm H\,I}$ using the measured column densities of \CII\ and \CIV, and the upper limits on the \ion{Si}{2} and \ion{Si}{4} columns as discussed in Section~4.1. Our primary interest is in the range of allowable ionization parameters that these data support, and the line-of-sight thicknesses that can be derived from these values, not the absorber metallicity, which we use only to constrain $N_{\rm H\,I}$.

\begin{figure}[!t]
\epsscale{1.00} \centering \plotone{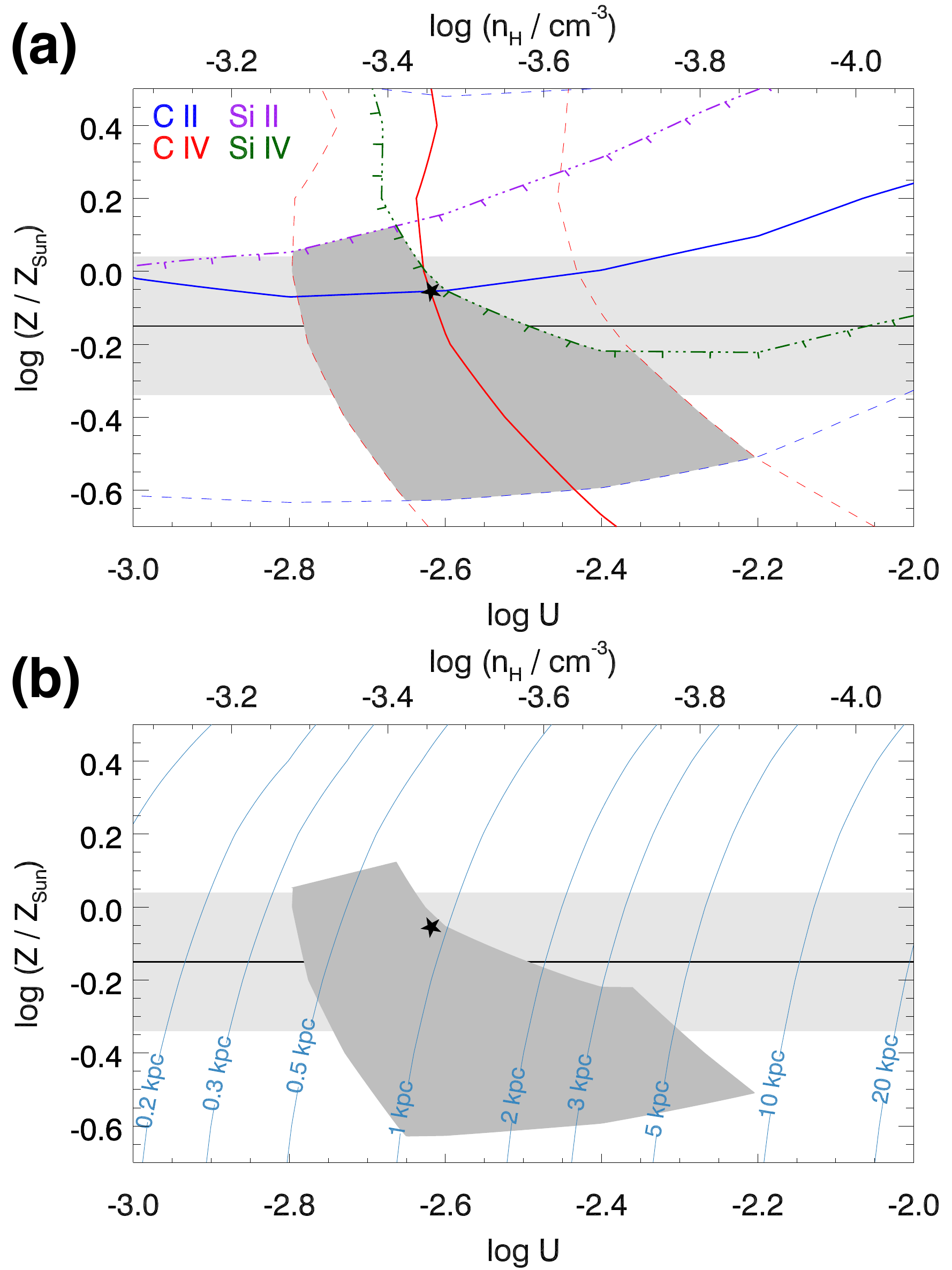}
\caption{{\it Top}: The range of {\it CLOUDY} models that satisfy all of the metal-line constraints for the $\sim1900$~\kms\ absorber toward J0107A with an assumed \HI\ column of $\log{N_{\rm H\,I}} = 15.5\pm0.5$ (see text). The allowed model region is shown shaded in gray with the various line constraints color-coded as in the Figure legend. The solid horizontal lines are the plausible metallicity limits set by the galaxy metallicity. {\it Bottom}: Contours of constant line-of-sight thickness, $t$, for the {\it CLOUDY} model in the top panel. The shaded region and metallicity indicators are the same as above.
\label{fig:cloudy}}
\end{figure}

The {\it CLOUDY} models associated with this revised column density for the $\sim1900$~\kms\ component in J0107A ($\log{N_{\rm H\,I}}=15.5\pm0.5$) are shown in Figure~5a. The metallicity of MCG~-01-04-005 is shown as a solid horizontal line with a light gray band indicating its $2\sigma$ uncertainty. The curved solid lines indicate the region of parameter space consistent with the measured \CII\ and \CIV\ column densities, with the flanking dashed lines showing the $1\sigma$ uncertainties in the measured columns modulo the very large uncertainty in the \HI\ column. The dot-dashed lines show the upper limits on the \ion{Si}{2} and \ion{Si}{4} columns with the tick marks pointing toward the allowable region of parameter space. The filled gray area shows the region of parameter space that is consistent with all of the constraints and the star indicates the fiducial solution where the measured \CII\ and \CIV\ columns agree exactly. 

From this model we estimate that $\log{U} = -2.6_{-0.2}^{+0.4}$ for this absorber; the large uncertainty reflects the large uncertainty in $N_{\rm H\,I}$. At the fiducial solution we find $T\sim11,000$~K, $n_{\rm H}\sim3\times10^{-4}~{\rm cm}^{-3}$, $f_{\rm H\,I}\sim0.3$\%, and $t\sim1$~kpc. The range of line-of-sight thicknesses allowed by our {\it CLOUDY} model is shown in Figure~5b, which displays the same shaded region and metallicity indicators as Figure~5a with contours of constant cloud thickness overlaid. From this plot we conclude that the indicative cloud thickness is $\approx0.5$--5~kpc.

Guided by the fiducial cloud thickness (1~kpc) and cloud density ($3\times10^{-4}~{\rm cm}^{-3}$) from the {\it CLOUDY} models and the cloud size limit ($>10$~kpc) on the sky from the common absorbers, a constant density ellipsoidal cloud with minimum sizes of 1 and 10~kpc in two of the three dimensions is suggested. Assuming 10~kpc for the third cloud dimension (second transverse dimension on the sky) finds a rough cloud mass limit of $\gtrsim3\times10^5~M_{\Sun}$. Allowing lower density and larger volume models as in Figure 5 does not change this limit appreciably. However, if our presumed \CIV\ identification for the absorption line detected in Q0107C is correct, the lower mass limit inferred for this spheroidal model is a factor of 4 larger. 

In summary, for all cases described in this Section, {\bf most lower limits on cloud mass} are in the range 0.3-$5\times10^6~M_{\Sun}$. If we had to rely on photo-ionization modeling alone, then the inferred mass would be about $100\times$ less than this lower limit by assuming that each cloud dimension is $\sim$ 1 kpc. Clearly, the physical limits derived from multiple sight lines are extremely valuable in estimating CGM cloud parameters. 

While these values make various assumptions to arrive at a lower mass limit, each of these limiting values for CGM cloud masses are better constrained than those found in either \citet{stocke13} or \citet{werk14} based solely on simple {\it CLOUDY} photo-ionization modeling. However, the current result supports the contention made by those authors that at least some CGM clouds have very large sizes and masses. And while we cannot, on the basis of current data, entirely rule out the contention of \citet{werk14} that the photo-ionized phase is a monolithic structure around each late-type galaxy, the limited size on the sky of $\sim10$~kpc for the lower velocity absorber argues weakly against that model.

\section{Conclusions}
\label{conclusion}

Using three neighboring QSO sight lines called the LBQS Triplet we have determined that two photo-ionized CGM gas clouds in the halo of the $\sim0.1\,L^*$ late-type galaxy MCG~-01-04-005 have physical extents on the sky of $\geq10$~kpc and total masses $\geq3\times10^5~M_{\Sun}$. Most size constraints available using common absorptions in the COS FUV spectra of the LBQS Triplet require minimum masses of $> 10^6~ M_{\Sun}$. Given that MCG~-01-04-005 is a typical, non-starburst low-luminosity galaxy, and that the impact parameters for these absorbers are typical for CGM absorbers at 0.4-0.5\,\Rvir, we argue that these results also are typical for photo-ionized CGM clouds in general. Therefore, these results support inferences by \citet{tumlinson11}, \citet{stocke13}, \citet{werk14}, and others that individual CGM clouds can be very massive and the ensemble of such clouds contains a significant baryon reservoir, comparable to the mass in a galaxy's stars or larger! The following results led us to these conclusions and represent the main observational findings of this paper.

\begin{enumerate}
\item{The absorption line data from the \hst/COS FUV spectra of the LBQS Triplet (J0107A,B,C) finds common absorption at $z=0.0063$ in two velocity components ($\sim$ 1830 and 1900 km s$^{-1}$) as follows: \lya\ in J0107A,B; \CIV\ in J0107A,C; \CII\ in J0107A,B; and upper limits for \CIV\ in J0107B, Si~II\, and Si~IV\ in J0107A,B. The common \lya\ and \CIV\ absorptions support minimum cloud sizes for both velocity components of $\geq10$~kpc. A more speculative detection of \CIV\ 1548 \AA\ in J0107C suggests an even larger size limit of $\geq23$~kpc. The measured values for the absorption line detections can be found in Table 2.}
\item{Using ground-based \Ha\ and space-based {\sl GALEX} imaging, we find that the star formation rate in MCG~-01-04-005 is small (between $\approx0.01$--$0.06~M_{\Sun}\,{\rm yr}^{-1}$) in the last 100 million years. We conclude that the galaxy is a normal, low luminosity late-type galaxy which has not had a significant starburst in the last $10^8$~years at least.}
\item{Using \CII\ and \CIV\ detections and limits on \ion{Si}{2} and \ion{Si}{4} at $\sim1900$~\kms\ in J01017A, single-phase photo-ionization models have been constructed imposing $Z_{\rm abs} \sim Z_{\rm gal}$ to limit the range of acceptable \HI\ columns (Section~4). These {\it CLOUDY} models find $\log{U} = -2.6_{-0.2}^{+0.4}$ for this absorber and derived physical quantities of $n_{\rm H}\sim3\times10^{-4}~{\rm cm}^{-3}$, $f_{\rm H\,I}\sim0.3$\%, and $t\sim1$~kpc. While this thickness ($t$) is smaller than the minimum size on the sky obtained from the common absorptions, the range of cloud thicknesses allowed by the {\it CLOUDY} models is consistent with an ellipsoidal cloud. However, these models are quite uncertain providing only indicative results on cloud thickness. The common absorptions in the three sight lines provide much better constraints requiring large cloud sizes and masses.}
\item{Since the common absorptions provide only firm lower limits on transverse clouds size, these observations cannot rule out the presence of a monolithic photoionzed phase gas around this galaxy. But the absence of \CIV\ absorption at $\sim$ 1830 km s$^{-1}$ in J0107C and the weakness of the \lya\ absorption at this velocity in J0107B argue weakly against that model \citep{werk14} by suggesting a cloud size not much larger than 10 kpc for that absorber.}
\end{enumerate}

\acknowledgments
Simon Morris and Neil Creighton are thanked for obtaining the excellent \hst/COS spectra used for this investigation. We thank the Apache Point Observatory (APO) for making available 3.5m observing time to obtain images and long-slit spectra of MGC~-01-04-005. The authors thank the National Science Foundation for support of this research at the University of Colorado through grant AST 1109117.

{\it Facilities:} \facility{HST (COS)}, \facility{GALEX}, \facility {APO 3.5m}

\end{document}

%% file: tab1.tex
\begin{deluxetable}{lccccc}

\tablecolumns{6}
\tablewidth{0pt}

\tablecaption{Summary of \hst/COS Observations
\label{tab:obs}}

\tablehead{ \colhead{Target} & \colhead{$z_{\rm em}$} & \colhead{Grating} & \colhead{$t_{\rm exp}$} & \colhead{S/N} & \colhead{Flux} }

\startdata
J0107A & 0.9570 & G130M & 28202 & 14 & $7\times10^{-16}$   \\
        &        & G160M & 44430 & 12 & $7\times10^{-16}$   \\  
J0107B & 0.9560 & G130M & 21239 & 16 & $1.0\times10^{-15}$ \\
        &        & G160M & 21149 & 11 & $1.3\times10^{-15}$ \\
J0107C & 0.7280 & G160M & 83433 & 10 & $3\times10^{-16}$
\enddata

\end{deluxetable}

%% file: all_species_table.tex
\begin{deluxetable*}{lcccccccc}

\tablecolumns{9}
\tablewidth{0pt}

\tablecaption{Absorption Table
\label{tab:lya}}

\tablehead{ \colhead{} & \colhead{} & \colhead{$\lambda_{0}$} & \colhead{$\lambda_{\rm obs}$} & \colhead{$cz$} & \colhead{$b$} & \colhead{\eqw} & \colhead{$\log{N}$\tablenotemark{a}} & \colhead{$\sigma$}\\
\colhead{Target} & \colhead{Species}& \colhead{(\AA)} & \colhead{(\AA)}  & \colhead{(\kms)} & \colhead{(\kms)} & \colhead{(m\AA)} & \colhead{(cm$^{-2}$)} & \colhead{} }
\startdata
J0107A\tablenotemark{b}        & Ly$\alpha$ &  1215.67  &  1223.09  &  $1830\pm10$ & $39\pm9$  & $213\pm117$ & $13.74\pm0.09$  &  19.0 \\
\nodata       & Ly$\alpha$ &  1215.67  &  1223.33  &  $1888\pm10$ & $40\pm5$  & $568\pm80$  & $14.75\pm0.20$  &  49.9 \\
\nodata       & \CIV\ 	   &  1548.20  &  1558.12  &  $1921\pm10$ & $50\pm10$ & $109\pm18$  & $13.47\pm0.08$  &  8.2  \\
\nodata       & \CIV\      &  1550.77  &  1560.71  &  1921 & 40 & 109  & 13.47  &  \nodata  \\
\nodata       & C II        &  1334.53  &  1342.96  &  $1893\pm6 $ & $12\pm10$ & $33\pm21 $  & $13.28\pm0.22$ &  6.6  \\
\nodata       & Si II       &  1260.42  &  1268.36  &  \nodata     & 20        & $<25$       & $<12.2$        &  $<3$ \\
\nodata       & Si IV       &  1402.77  &  1411.61  &  \nodata     & 20        & $<22$       & $<12.7$        &  $<3$ \\
J0107B\tablenotemark{c}   & Ly$\alpha$ &  1215.67  &  1223.12 &  $1836\pm75$ &  $4\pm10$  &   34    &   13.16 &  6.2  \\
\nodata	                  & Ly$\alpha$ &  1215.67  &  1223.41  &  $1908\pm10$ & $17\pm12$ &   354   &   15.84 &  45.0 \\
\nodata       & \CIV\ & 1548.20  &  \nodata  &  \nodata     & \nodata   &    $<57$     &    $<13.43$          &  $<3$ \\
\nodata	      & \CIV\ & 1550.77  &  \nodata  &  \nodata     & \nodata   &    $<57$     &    $<13.45$          &  $<3$\\  
\nodata       & C II   & 1334.53  &  1343.17  &  $1939\pm15$ & $27\pm23$ &    $24\pm22$ &    $13.10\pm0.24$   &  $\sim3$ \\
\nodata       & Si II  & 1260.42  &  1268.36  &  \nodata     & 20        &    $<23$     &    $<12.2$          &  $<3$    \\
\nodata       & Si IV  & 1402.77  &  1411.62  &  $1891\pm10$ & $21\pm10$ &    $36\pm22$ &    $12.95\pm0.20$   &  3.2   \\
J0107C\tablenotemark{d}  & \CIV\ &  1548.20  &  1557.97  &  $1893\pm7$  & $34\pm11$ & $96\pm44$  & $13.43\pm0.15$  & 8.8   \\
\nodata		             & \CIV\ &  1550.77  &  \nodata  &  1893        &  34       &  96        &  13.43          & \nodata
\enddata
\tablenotetext{a}{The errors on $\log{N}$ represent formal fitting errors from $\chi^{2}$ minimization. We have reason to believe the errors on $\log{N_{\rm H\,I}}$ are actually much larger, as discussed in Section 4.}
\tablenotetext{b}{The weak line of the C\,IV doublet is assumed to be blended with an unknown weak line, as Voigt fitting results in a column density that is too high given known line ratios for C\,IV. We adopt the strong doublet line values for the 1550~\AA\ line as a reasonable proxy.}
\tablenotetext{c}{Since the C\,IV values for this sight line are upper limits, the observed wavelengths are based on the host galaxy redshift and assumed to be 1558.07 \AA~ and 1560.50 \AA. These have been used as the profile centers in Figure 2.}
\tablenotetext{d}{The weak line of the C\,IV doublet is highly blended with unidentified lines and thus no measurement was made, so we follow the same procedure as footnote b.}
\end{deluxetable*}